\documentclass[twocolumn,twocolappendix]{aastex701}
\usepackage{booktabs}
\usepackage{longtable}
\usepackage{amsmath}
\usepackage{threeparttable}

\received{April 12, 2026}
\revised{June 10, 2026}
\accepted{June 10, 2026}
\submitjournal{ApJ}

\begin{document}

\title{Chandra X-Ray Imaging and Spatially Resolved Spectroscopy of SN~1987A: \\
Energy-Dependent Morphology of the Equatorial Ring}

\author[0000-0002-5809-3516]{Yusuke Sakai}
\affiliation{Department of Physics, Rikkyo University, Toshima-ku, Tokyo, 171-8501, Japan}
\affiliation{National Institute of Advanced Industrial Science and Technology, Tsukuba, 305-8560, Japan}
\email[show]{sakai.yusuke.d@rikkyo.ac.jp}

\author[0000-0003-4808-893X]{Shinya Yamada}
\affiliation{Department of Physics, Rikkyo University, Toshima-ku, Tokyo, 171-8501, Japan}
\email{hoge@gmail.com}

\author[0000-0002-0018-0369]{Koji Mori}
\affiliation{Faculty of Engineering, University of Miyazaki, Miyazaki 889-2192, Japan}
\email{hoge@gmail.com}

\author[0000-0002-8152-6172]{Hiromasa Suzuki}
\affiliation{Faculty of Engineering, University of Miyazaki, Miyazaki 889-2192, Japan}
\email{hoge@gmail.com}

\author[0000-0002-4037-1346]{Haruka Sakemi}
\affiliation{Graduate School of Sciences and Technology for Innovation, Yamaguchi University, 1677-1 Yoshida, Yamaguchi, Yamaguchi 753-0841, Japan}
\affiliation{Nobeyama Radio Observatory, National Astronomical Observatory of Japan (NAOJ), National Institutes of Natural Sciences (NINS), 462-2, Nobeyama, Minamimaki, Minamisaku, Nagano 384-1305, Japan}
\email{hoge@gmail.com}

\author[]{Tsukasa Matsushima}
\affiliation{Faculty of Engineering, University of Miyazaki, Miyazaki 889-2192, Japan}
\email{hoge@gmail.com}

\author[0009-0009-0927-0772]{Shintaro Kaneko}
\affiliation{Department of Physics, Rikkyo University, Toshima-ku, Tokyo, 171-8501, Japan}
\email{hoge@gmail.com}

\author[0009-0003-0653-2913]{Kai Matsunaga}
\affiliation{Department of Physics, Kyoto University, Kitashirakawa Oiwakecho, Sakyo-ku, Kyoto 606-8502, Japan}
\email{hoge@gmail.com}

\author[0000-0001-7773-9266]{Shogo B. Kobayashi}
\affiliation{Department of Physics, Rikkyo University, Toshima-ku, Tokyo, 171-8501, Japan}
\email{hoge@gmail.com}

\author[]{Haruto Aoki}
\affiliation{Department of Physics, Rikkyo University, Toshima-ku, Tokyo, 171-8501, Japan}
\email{hoge@gmail.com}

\author[0000-0001-9267-1693]{Toshiki Sato}
\affiliation{School of Science and Technology, Meiji University, 1-1-1 Higashi-Mita, Tama-ku, Kawasaki, Kanagawa, 214-8571, Japan}
\email{hoge@gmail.com}

\begin{abstract}
We present a systematic imaging and spatially resolved spectral study of SN 1987A using Chandra observations obtained between 1999 and 2025. By combining multiepoch ACIS and HETG data, we investigate the long-term evolution of the remnant in both the soft and hard X-ray bands. To characterize the radial structure, we model the projected emission with a torus profile and derive its radius and width on the image plane. We find an energy dependence in the ring morphology: while the soft and hard bands exhibit similar structures at early epochs, the soft-band emission becomes systematically broader than the hard-band emission after the early 2010s. Furthermore, when considering the radius and width together, the soft-band emission shows an inward extension, suggesting an increasing contribution from interior and/or high-latitude emission components. The flux evolution of the Fe~K line is consistent with previous XMM-Newton results, and we detect its presence already in earlier epochs ($\sim$2007--2009) using combined Chandra spectra. Spatially resolved analysis further indicates that the Fe~K emission is enhanced in the eastern region. These results provide a unified view of the long-term morphological and spectral evolution of SN~1987A and highlight the emergence of energy-dependent radial structure as a key feature in its late-time evolution.
\end{abstract}


\section{Introduction}

SN~1987A, located in the Large Magellanic Cloud at a distance of
$\sim$50~kpc, is one of the best-studied core-collapse supernovae in the
modern era. Since its explosion on 1987 February 23, it has provided a
unique opportunity to follow, on human timescales, the transition from a
supernova to a young supernova remnant
\citep{West_1987,Arnett_1989,McCray_1993,McCray_2016}. Owing to its
proximity and the extensive monitoring carried out over nearly four
decades across the electromagnetic spectrum, SN~1987A has become a
benchmark object for studying the interaction between the explosion
ejecta and the surrounding circumstellar material (CSM).

A defining feature of SN~1987A is its structured CSM, most notably the
triple-ring system revealed by optical observations, consisting of a
bright inner equatorial ring (ER) and two fainter outer rings
\citep{Burrows_1995}. The CSM is thought to consist of a dense, clumpy
ER embedded in a more diffuse H\,\textsc{ii} region
\citep{Chevalier_1995,Sugerman_2005}. As the blast wave has propagated
through this environment, its interaction with the dense ER has produced
the strong rise of the X-ray emission and the development of the
characteristic ring-like X-ray morphology
\citep{Burrows_2000,Park_2002,Park_2004,Park_2005}. 
This ring-like emission is not azimuthally uniform; Chandra
observations have shown pronounced east--west asymmetry in the
X-ray emission, and subsequent imaging and spectroscopic studies
revealed that the two sides evolve differently
\citep{Michael_2002,Zhekov_2009,Frank_2016,Ravi_2024}.
For much of the Chandra era, the X-ray emission from SN~1987A has
therefore been interpreted primarily as originating from shock-heated
ER material while retaining information on the asymmetric interaction
between the blast wave and the structured CSM.

At the same time, the X-ray morphology is not identical across energy
bands. Early Chandra studies showed that the soft-band emission
is more closely associated with dense ER protrusions, whereas the harder
emission traces faster shocks and has a different spatial distribution
\citep{Park_2004,Park_2006,Park_2007}. More recent observations indicate
that this trend has evolved, with the soft-band emission extending toward smaller radii, while the hard-band emission extends to larger
radii \citep{Ravi_2024}.

SN~1987A is now entering a new stage of its evolution. Long-term
Chandra monitoring shows that the soft X-ray light curve has
flattened \citep{Frank_2016}. Subsequent studies report a decline in the
soft X-ray flux, continued evolution in the hard band, and changes in
the expansion behavior \citep{Sun_2021,Sun_2025, Ravi_2024}. Optical observations show
fading of the ER, brightening and structural evolution of the ejecta,
and new emission outside the ER \citep{Larsson_2019}. JWST observations
further reveal complex shock structures both in and around the ER and at
higher latitudes \citep{Larsson_2023}.

This emerging picture is broadly consistent with recent
three-dimensional hydrodynamic (HD) and magnetohydrodynamic (MHD)
simulations, which predict that, about 30--35 yr after the
explosion, the X-ray emission should show an increasing contribution
from shocked ejecta while still retaining information on the explosion
geometry and the progenitor environment
\citep{Orlando_2015,Orlando_2019,Orlando_2020,Orlando_2025}.

The Fe~K emission around 6.7~keV provides an important probe of the hottest plasma in the
remnant. This component has been reported in XMM-Newton
observations since 2010 \citep{Sturm_2010,Maggi_2012,Sun_2021}, and more
recent Chandra studies suggested that it became clearly
detectable after 2018 and may be enhanced on the eastern side of the
ring \citep{Ravi_2024}. However, limited photon statistics in the hard
band have so far made it difficult to investigate the spatial evolution
of the Fe~K-emitting region in detail.

A systematic investigation of these long-term morphological and spectral trends
requires combining multiepoch Chandra observations. However, this is
not straightforward for SN~1987A, because its subarcsecond structure
makes merged images sensitive to small relative astrometric offsets.
In addition, reliable astrometric registration based on reference sources
is not always feasible due to the limited field of view. The available
dataset includes observations obtained over more than two decades,
making a consistent treatment of relative alignment essential.

In this paper, we present a systematic imaging and spatially resolved
spectral study of SN~1987A using Chandra observations obtained
between 1999 and 2025. By combining nongrating (NONE) ACIS data and HETG
zeroth-order images over multiyear intervals, we investigate the
long-term evolution of the remnant in the full, soft, and hard X-ray
bands. Our primary goal is to quantify, in a uniform manner, how the projected
radius and width of the X-ray-emitting ring evolve with time and to
determine whether the soft and hard bands show distinct radial
morphological evolution as SN~1987A moves beyond the classical
ER-dominated phase. In particular, we focus on whether the energy
dependence of the torus width and the inward extension of the soft-band
emission can be identified from the long-term Chandra dataset. We also examine the Fe~K emission using combined and spatially resolved spectra to investigate its long-term and spatial evolution.

\section{Data Reduction and Analysis}
\label{sec:data}

\subsection{Chandra dataset}

We analyzed archival Chandra observations of SN~1987A obtained between 1999 and 2025. 
The dataset consists of ACIS-S observations, including both NONE and zeroth-order HETG data. 
In total, 21 NONE observations ($\sim$660~ks) and 79 HETG observations ($\sim$2560~ks) were used.
The data reduction was performed using CIAO \citep{Fruscione_2006} version~4.17 with CALDB version~4.12.2.
Standard data reprocessing was carried out using \texttt{chandra\_repro}.
 
As discussed in \citet{Helder_2013}, \citet{Frank_2016}, and \citet{Ravi_2024}, the ACIS observing configurations were modified several times after 2008 to mitigate photon pileup caused by the increasing brightness of SN~1987A, including changes in frame time and the insertion of HETG for many observations. 
For the present imaging analysis, we checked the count rate per frame
per ACIS detection cell ($3 \times 3$ native pixels) in the brightest regions of
both the ACIS/NONE and HETG zeroth-order images. The peak values were typically below $\sim0.2$
counts frame$^{-1}$ detection cell$^{-1}$,
corresponding to estimated pileup fractions of roughly
$\lesssim10\%$. We note that several bright epochs between 2002 and 2004 reached higher values.

To improve photon statistics while preserving temporal evolution, the observations were grouped into 4 yr bins for subsequent analysis, which provides a balance between temporal resolution and sufficient photon statistics for reliable spectral fitting. HETG and NONE data were treated separately due to differences in instrumental configuration and systematic characteristics. The resulting binning scheme is summarized in Table~\ref{tab:merge_obs_info}.
Throughout this paper, uncertainties are quoted at the 1$\sigma$ confidence level unless otherwise noted.

\begin{table}[ht!]
\centering
\caption{Summary of 4 yr merged Chandra observations of SN~1987A.}
\label{tab:merge_obs_info}
\begin{tabular}{ccrrr}
\toprule
Year$^{*}$ & Grating & $N_{\rm obs}^{\dagger}$ & Age$^{\ddagger}$ & Exposure$^{\S}$ \\
 & &  & (days) & (ks) \\
\midrule
1999--1999 & HETG & 2 & 4608 & 116.2 \\
2000--2003 & NONE & 7 & 5422 & 313.1 \\
2004--2007 & NONE & 10 & 6738 & 326.4 \\
2007--2009 & HETG & 27 & 7734 & 645.1 \\
2011--2014 & HETG & 14 & 9288 & 623.1 \\
2015--2018 & HETG & 20 & 11183 & 582.2 \\
2019--2022 & HETG & 9 & 12454 & 343.6 \\
2023--2025 & HETG & 7 & 13727 & 253.3 \\
\bottomrule
\end{tabular}
\begin{tablenotes}
\item \textit{Notes.}
\item $^{*}$ In the 4 yr binning, NONE observations in the 2008--2011 interval ($\sim$20~ks) were not included as contemporaneous HETG data provide substantially deeper coverage.
\item $^{\dagger}$ Number of individual ObsIDs included in each combined dataset. For the full list of ObsIDs, see Table~\ref{tab:obslog}.
\item $^{\ddagger}$ The age is defined as the exposure-time-weighted mean of the observation ages of individual ObsIDs in each bin.
\item $^{\S}$ The exposure is defined as the total effective exposure time summed over all ObsIDs in each bin.
\end{tablenotes}
\end{table}

\begin{figure*}[ht!]
\centering
 \includegraphics[width=1.0\linewidth]{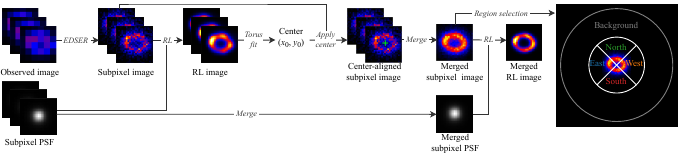}
\caption{Flowchart of the image and spectral analysis procedures for the binned RL deconvolution.
For each observation, the original Chandra image (bin = 1) is processed with EDSER to reconstruct a higher-resolution image (bin = 1/4), and a PSF with the same binning is simulated for RL deconvolution.
The deconvolved image in the 1.5--7.0~keV band is fitted with a torus model to determine the center position.
This procedure is applied independently to all observations, after which the images are aligned and merged.
The PSFs are combined using exposure-weighted averaging, and the merged image is deconvolved using RL with the corresponding merged PSF.
The same procedure is applied to other energy bands using the center position determined from the 1.5--7.0~keV band.
Regions for spectral extraction are shown on the merged image.}
\label{sn1987a_flowchart}
\end{figure*}

\subsection{Overview of the merge strategy}
\label{subsec:merge_overview}

Accurate relative alignment is essential for merging multi-epoch images. 
In many Chandra studies, this is achieved using WCS corrections based on nearby point sources (e.g., \citealt{Sato_2018, Suzuki_2023, Sakai_2024}). 
However, this approach is not applicable to the present dataset. 
Most ACIS observations were taken in subarray mode, resulting in a limited field of view, and the HETG observations are dominated by dispersed grating arms. 
As a result, it is difficult to identify consistent point sources for astrometric correction.

We therefore adopt an internal alignment strategy using the remnant itself as a reference. 
The center of the remnant is determined for each observation using a combination of subpixel reconstruction and deconvolution to enhance the ring structure.
The original exposure-corrected images are then aligned and merged to maximize photon statistics.
Finally, RL deconvolution is applied to the merged image.
A schematic overview of this procedure is shown in Figure~\ref{sn1987a_flowchart}. 
The detailed data reduction and analysis steps are described in the following subsections.

\subsection{Center determination}

We determine the reference center for each observation.
The ACIS event data were processed with energy-dependent subpixel event repositioning (EDSER; \citealt{Li_2004}), based on earlier event repositioning techniques \citep{Mori_2001,Tsunemi_2001,  Li_2003}. 
We adopt a subpixel scale of 1/4 pixel.

We use images in the 1.5--7.0~keV band, constructed with \texttt{fluximage}, where
the ring structure is relatively well defined. For each ObsID, a monochromatic (2.8~keV) point-spread function (PSF) is simulated using MARX \citep{davis2012raytracing} version~5.5.3 with the same spatial resolution as the subpixel images. RL deconvolution is then applied with 10
iterations to mitigate PSF blurring while preserving morphological stability.
The resulting image is fitted with an elliptical torus model, and the best-fit
center is adopted as the reference position.

\begin{figure*}[ht!]
 \includegraphics[width=1.0\linewidth]{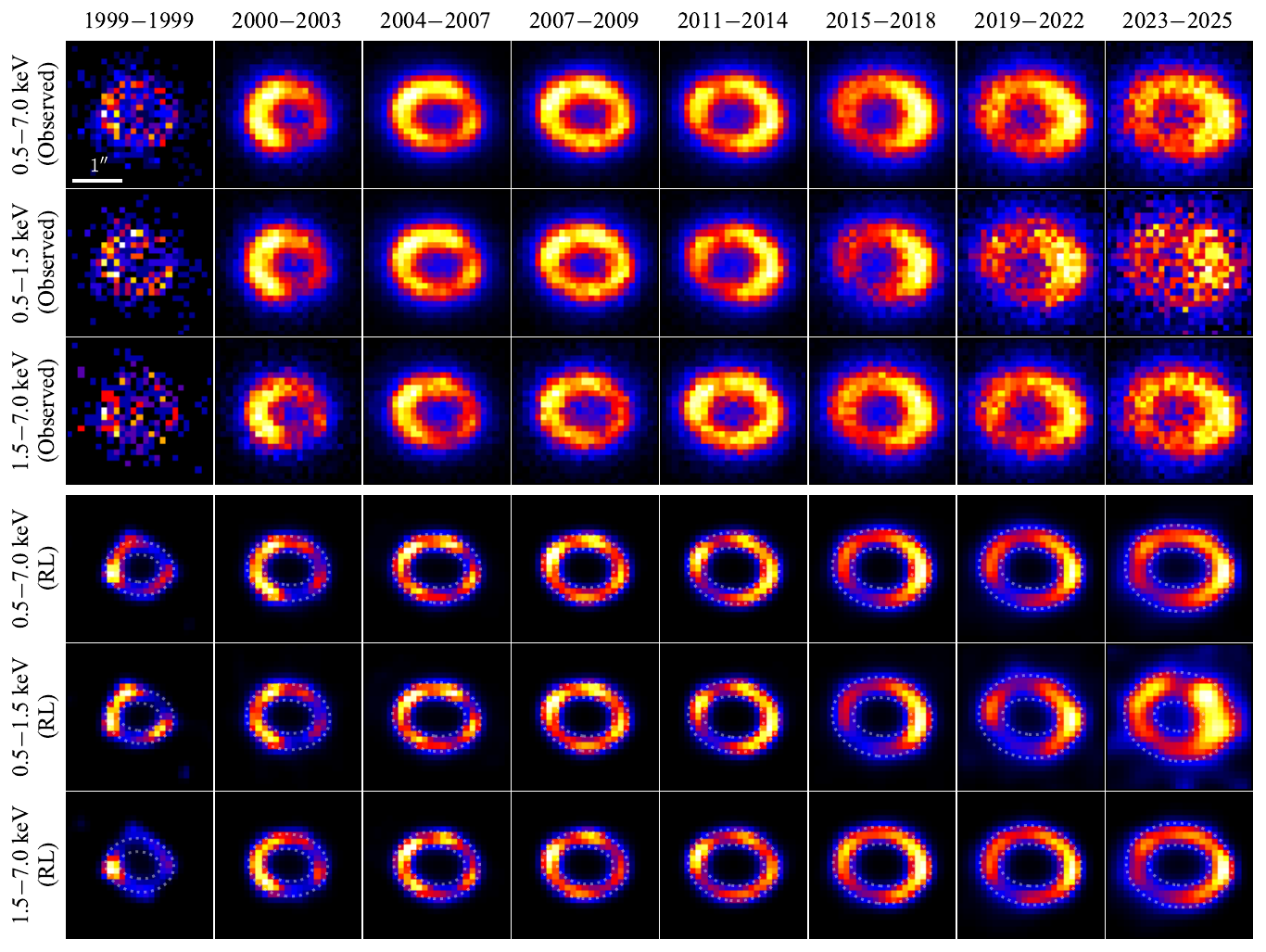}
\caption{The 4 yr combined X-ray images showing the temporal evolution of the torus structure.
The top three rows present the observed images reconstructed with EDSER (1/4 pixel) in the 0.5--7.0, 0.5--1.5, and 1.5--7.0~keV bands.
The bottom rows show the corresponding images after RL deconvolution.
Each column corresponds to a 4 yr epoch (see Table~\ref{tab:merge_obs_info}).
The color scale in each panel is linearly scaled between zero and the maximum value.
The dotted rings in the RL images indicate the torus boundaries ($\pm \sigma_r$) derived from the fitted torus model.}
 \label{sn1987a_torus_image_obs_rl_iter_30_bin_0.25_yearbin_4}
\end{figure*}

\subsection{Imaging analysis}

The images were reconstructed with EDSER at a 1/4 pixel scale for
each observation and combined in three energy bands
($0.5$--$7.0$, $0.5$--$1.5$, and $1.5$--$7.0$~keV). The standard CIAO image-merging scripts were not directly used. Instead, the subpixel-reconstructed images and
corresponding exposure maps were combined through direct
exposure-corrected co-addition. 

The corresponding PSFs were simulated using MARX for each observation
at representative energies of $2.2$, $1.1$,
and $2.8$~keV for the three energy bands, respectively, where
the representative energies were chosen based on the average count
distribution in each band. The PSFs were then merged using
exposure-weighted combinations. RL deconvolution was applied to the
merged images using the corresponding merged PSFs. The number of
iterations was set to 30, balancing resolution enhancement and noise
amplification.

\subsection{Spectral analysis}

For the spectral analysis, source regions were defined as circular apertures with radii $0''$--$3''$ centered on the reference position, while background regions were defined as annuli with radii $6''$--$12''$. To investigate spatial dependence, the source region was further divided into four sectors (East, West, North, and South), each spanning $\pm45^\circ$, in addition to a full circular region (All), resulting in five regions (Figure~\ref{sn1987a_flowchart}). For each ObsID, spectra were extracted from these regions and combined using \texttt{combine\_spectra}, with the corresponding response matrix files (RMFs) and ancillary response files (ARFs) combined simultaneously. 

Spectral fitting was performed using XSPEC \citep{Arnaud_1996} version~12.15.1 with the W-statistic \citep{Cash_1979,Wachter_1979}, which is appropriate for Poisson-distributed source and background spectra.
In XSPEC, the W-statistic is internally applied when the \texttt{cstat} option is used together with source and background spectra.
The spectra were analyzed in the 5.0--8.0~keV band to focus on the Fe~K emission.

\subsection{Contemporaneous Multiwavelength Data}

To place the X-ray morphology in a broader multiwavelength context, we used radio, optical, and near-infrared data obtained in 2022.
The radio data were obtained with ATCA on 2022 February 2 and reduced using CASA \citep{Team_2022} version 6.7.2. Two frequency ranges, $8461$--$8564$ and $8845$--$8948$~MHz were combined to produce a continuum image using the \texttt{tclean} task with multifrequency synthesis. The rms noise level, estimated from source-free outer regions, is $\sim9\times10^{-4}$~Jy~beam$^{-1}$.

The near-infrared data were obtained with JWST (F444W) on 2022 September 2 \citep{Matsuura_2024} and retrieved from the Mikulski Archive for Space Telescopes (MAST). The optical data were obtained with the Hubble Space Telescope (HST; F625W) on 2022 September 5 \citep{Tegkelidis_2024} and retrieved from MAST. For the X-ray comparison, we used the Chandra observation taken on 2022 September 21 (ObsID 25514) in the 0.5--1.5 and 1.5--7.0~keV bands, where images deconvolved with 10 iterations of the RL method were used. 

\section{Results}

\subsection{Imaging analysis}
\label{sec:imaging}
Merged images for each 4 yr bin are shown in Figure~\ref{sn1987a_torus_image_obs_rl_iter_30_bin_0.25_yearbin_4}, together with the corresponding EDSER-reconstructed images (top rows) and RL-deconvolved images (bottom rows). 
The images show an asymmetric structure of the ER. In the early epochs, the eastern side is brighter, whereas the western side becomes dominant after 2007--2009 in the 0.5--7.0~keV band. This behavior is consistent with previous studies \citep{Frank_2016}. 

\begin{figure}[ht!]
 \includegraphics[width=1.0\linewidth]{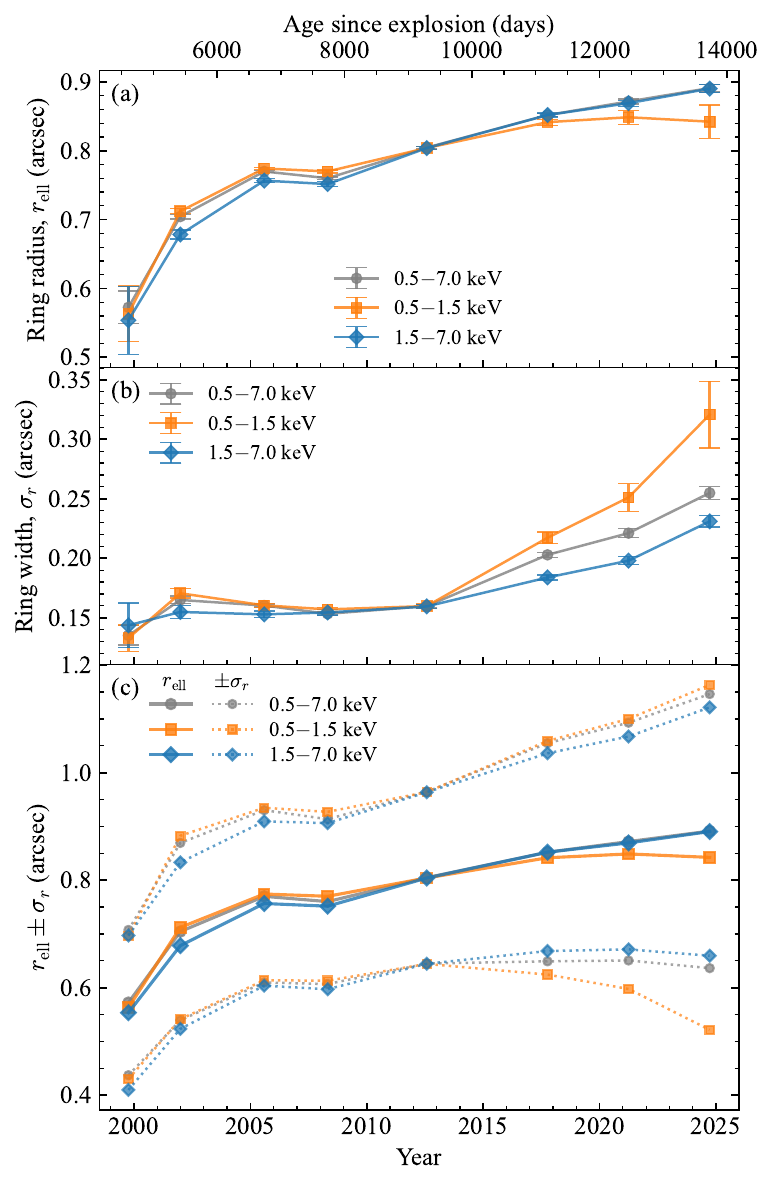}
\caption{Temporal evolution of the torus radius ($r_{\mathrm{ell}}$) and width ($\sigma_r$) derived from RL-deconvolved images.
Results are shown for the 0.5--7.0~keV (gray), 0.5--1.5~keV (orange), and 1.5--7.0~keV (blue) bands.
(a) Torus radius $r_{\mathrm{ell}}$.
(b) Torus width $\sigma_r$.
(c) Combined representation showing $r_{\mathrm{ell}} \pm \sigma_r$.
The plotted ages correspond to the exposure-time-weighted mean ages of
the observations in each merged dataset, as listed in
Table~\ref{tab:merge_obs_info}.}
 \label{sn1987a_torus_param_bin_0.25_yearbin_4}
\end{figure}

To quantify the radial position and width of the ring structure, we performed torus model fitting on the RL-deconvolved images.  
We adopt a model following the two-dimensional projected torus geometry described in \citet{Racusin_2009}.
For a given inclination and position angle ($i = 42\fdg85$ and $\mathrm{PA} = -6\fdg24$; \citealt{Tegkelidis_2024}), 
the projected structure appears as an ellipse, and the radial profile is therefore evaluated in this elliptical coordinate system. 
We note that the radius and width derived here are evaluated on the projected two-dimensional image and thus represent projected quantities. 
While \citet{Racusin_2009} deprojected the image to a circular geometry, we instead model the ring directly in the projected plane to avoid introducing potential systematic uncertainties associated with geometric transformations of the image.
Previous studies have employed more complex models, such as those including four-lobe structures \citep[e.g.,][]{Racusin_2009, Frank_2016}. 
In this work, we adopt a simple torus model with a uniform emissivity distribution to ensure a consistent analysis across all datasets (see Appendix~\ref{app:torus_model} for the model formulation).

The best-fit model is overlaid in Figure~\ref{sn1987a_torus_image_obs_rl_iter_30_bin_0.25_yearbin_4}, 
showing the ring structure and its radial extent, and is in good agreement with the observed morphology. 
The temporal evolution of the ring major-axis radius $r_{\mathrm{ell}}$ and width $\sigma_r$ 
is shown in Figure~\ref{sn1987a_torus_param_bin_0.25_yearbin_4}. 
The evolution of the ring radius (Figure~\ref{sn1987a_torus_param_bin_0.25_yearbin_4}(a)) is consistent with previous measurements 
\citep[e.g.,][]{Racusin_2009, Frank_2016, Ravi_2024}, supporting the validity of our analysis. 
We note that the ring radius appears slightly larger during 2004--2007 (NONE) 
than in 2007--2009 (HETG). This difference has been suggested to arise from instrumental effects, 
specifically that NONE observations tend to yield slightly larger radii than HETG observations 
\citep{Ravi_2024}.

The temporal evolution of the ring width has been reported previously \citep{Ravi_2024}, 
and our measurements (Figure~\ref{sn1987a_torus_param_bin_0.25_yearbin_4}(b)) 
are consistent with this trend. 
Building on these previous measurements, the main new result of our
energy-resolved imaging analysis is that the width evolution is
energy dependent. Up to $\sim$2012, the ring width is similar across all
energy bands, whereas after this epoch the soft (0.5--1.5~keV) band
becomes systematically broader than the hard (1.5--7.0~keV) band.
This trend is also suggested by the torus fitting results for
individual observations
(Appendix~\ref{appendix:individual_torus}), indicating that it is not
driven solely by the merging procedure, and is further supported by an
independent radial profile analysis designed to mitigate model
dependence (Appendix~\ref{appendix:radial_profile}).

The combination of the ring radius and width is shown in 
Figure~\ref{sn1987a_torus_param_bin_0.25_yearbin_4}(c). 
From this figure, it is evident that in the 2020s the width in the soft band becomes larger, while the characteristic radius slightly decreases, 
resulting in an inward extension of the emission. 
In particular, the inner boundary of the distribution in the 2020s reaches down to 
$r_{\mathrm{ell}} \sim 0\farcs5$, comparable to the radius observed around the year 2000. 
In contrast, the outer boundary of the ring remains comparable between the soft and hard bands, indicating that the energy dependence arises primarily from the inner region.

\begin{table*}[ht!]
\caption{Fit parameters of the \texttt{tbabs*(powerlaw + gaussian)} model in the 5.0--8.0 keV band.}
\label{tab:powerlaw_gauss}
\begin{tabular}{llccccc}
\toprule
Year & Region & Line flux & Line energy & Power-law flux$^\mathrm{a}$ & Photon index & $W$-stat/d.o.f. \\
 &  & (10$^{-6}$ ph cm$^{-2}$ s$^{-1}$) & (keV) & (10$^{-13}$ erg cm$^{-2}$ s$^{-1}$) &  & \\
\midrule
2000--2003 & All & $\mathrm{N/C}^\mathrm{b}$ & $\mathrm{N/C}^\mathrm{b}$ & $0.54^{+0.04}_{-0.04}$ & $1.66^{+0.54}_{-0.50}$ & 42/47 \\
2000--2003 & East & $\mathrm{N/C}^\mathrm{b}$ & $\mathrm{N/C}^\mathrm{b}$ & $0.22^{+0.03}_{-0.02}$ & $\mathrm{N/C}^\mathrm{b}$ & 18/16 \\
2000--2003 & West & $\mathrm{N/C}^\mathrm{b}$ & $\mathrm{N/C}^\mathrm{b}$ & $0.09^{+0.02}_{-0.02}$ & $5.14^{+2.27}_{-1.81}$ & 11/7 \\
2000--2003 & North & $\mathrm{N/C}^\mathrm{b}$ & $\mathrm{N/C}^\mathrm{b}$ & $0.10^{+0.02}_{-0.02}$ & $0.78^{+1.55}_{-1.36}$ & 3/6 \\
2000--2003 & South & $\mathrm{N/C}^\mathrm{b}$ & $\mathrm{N/C}^\mathrm{b}$ & $0.12^{+0.02}_{-0.02}$ & $2.17^{+1.24}_{-1.20}$ & 13/10 \\
\midrule
2004--2007 & All & $\mathrm{N/C}^\mathrm{b}$ & $\mathrm{N/C}^\mathrm{b}$ & $1.19^{+0.06}_{-0.06}$ & $2.76^{+0.37}_{-0.35}$ & 105/94 \\
2004--2007 & East & $0.16^{+0.13}_{-0.11}$ & $6.63^{+0.08}_{-0.08}$ & $0.44^{+0.04}_{-0.04}$ & $1.93^{+0.65}_{-0.63}$ & 41/40 \\
2004--2007 & West & $\mathrm{N/C}^\mathrm{b}$ & $\mathrm{N/C}^\mathrm{b}$ & $0.25^{+0.03}_{-0.03}$ & $3.37^{+0.87}_{-0.87}$ & 20/27 \\
2004--2007 & North & $\mathrm{N/C}^\mathrm{b}$ & $\mathrm{N/C}^\mathrm{b}$ & $0.24^{+0.03}_{-0.03}$ & $4.35^{+0.93}_{-0.81}$ & 20/26 \\
2004--2007 & South & $\mathrm{N/C}^\mathrm{b}$ & $\mathrm{N/C}^\mathrm{b}$ & $\mathrm{N/C}^\mathrm{b}$ & $2.67^{+0.84}_{-0.84}$ & 26/23 \\
\midrule
2007--2009 & All & $0.75^{+0.24}_{-0.22}$ & $6.62^{+0.03}_{-0.03}$ & $1.99^{+0.07}_{-0.07}$ & $2.33^{+0.27}_{-0.26}$ & 132/136 \\
2007--2009 & East & $0.31^{+0.14}_{-0.13}$ & $6.61^{+0.04}_{-0.04}$ & $0.58^{+0.04}_{-0.04}$ & $2.45^{+0.50}_{-0.49}$ & 60/59 \\
2007--2009 & West & $\mathrm{N/C}^\mathrm{b}$ & $\mathrm{N/C}^\mathrm{b}$ & $0.46^{+0.04}_{-0.03}$ & $2.66^{+0.55}_{-0.56}$ & 51/46 \\
2007--2009 & North & $0.18^{+0.13}_{-0.11}$ & $6.72^{+0.07}_{-0.07}$ & $0.49^{+0.04}_{-0.03}$ & $2.75^{+0.55}_{-0.54}$ & 44/47 \\
2007--2009 & South & $0.24^{+0.12}_{-0.11}$ & $6.56^{+0.05}_{-0.04}$ & $0.44^{+0.04}_{-0.03}$ & $1.86^{+0.57}_{-0.54}$ & 32/44 \\
\midrule
2011--2014 & All & $1.97^{+0.34}_{-0.33}$ & $6.67^{+0.02}_{-0.01}$ & $3.60^{+0.10}_{-0.09}$ & $3.07^{+0.20}_{-0.19}$ & 142/158 \\
2011--2014 & East & $0.78^{+0.21}_{-0.19}$ & $6.71^{+0.03}_{-0.03}$ & $1.11^{+0.05}_{-0.05}$ & $2.87^{+0.36}_{-0.36}$ & 80/95 \\
2011--2014 & West & $0.34^{+0.17}_{-0.15}$ & $6.62^{+0.04}_{-0.04}$ & $0.98^{+0.05}_{-0.05}$ & $2.94^{+0.38}_{-0.37}$ & 91/87 \\
2011--2014 & North & $0.58^{+0.18}_{-0.16}$ & $6.65^{+0.03}_{-0.03}$ & $0.80^{+0.05}_{-0.04}$ & $3.31^{+0.44}_{-0.43}$ & 51/76 \\
2011--2014 & South & $0.39^{+0.15}_{-0.14}$ & $6.67^{+0.04}_{-0.03}$ & $0.70^{+0.04}_{-0.04}$ & $3.40^{+0.48}_{-0.47}$ & 57/67 \\
\midrule
2015--2018 & All & $3.20^{+0.45}_{-0.43}$ & $6.65^{+0.02}_{-0.01}$ & $5.53^{+0.12}_{-0.12}$ & $2.58^{+0.16}_{-0.16}$ & 210/177 \\
2015--2018 & East & $0.96^{+0.24}_{-0.22}$ & $6.62^{+0.02}_{-0.02}$ & $1.59^{+0.07}_{-0.07}$ & $2.65^{+0.31}_{-0.30}$ & 106/115 \\
2015--2018 & West & $1.13^{+0.25}_{-0.24}$ & $6.68^{+0.03}_{-0.02}$ & $1.48^{+0.07}_{-0.06}$ & $2.80^{+0.32}_{-0.32}$ & 125/111 \\
2015--2018 & North & $0.59^{+0.21}_{-0.20}$ & $6.72^{+0.04}_{-0.03}$ & $1.29^{+0.06}_{-0.06}$ & $2.27^{+0.35}_{-0.34}$ & 85/102 \\
2015--2018 & South & $0.67^{+0.20}_{-0.19}$ & $6.63^{+0.03}_{-0.02}$ & $1.17^{+0.06}_{-0.06}$ & $2.48^{+0.36}_{-0.35}$ & 64/93 \\
\midrule
2019--2022 & All & $5.15^{+0.69}_{-0.66}$ & $6.66^{+0.01}_{-0.01}$ & $6.53^{+0.18}_{-0.17}$ & $2.64^{+0.20}_{-0.20}$ & 156/160 \\
2019--2022 & East & $2.36^{+0.41}_{-0.41}$ & $6.68^{+0.01}_{-0.02}$ & $1.95^{+0.10}_{-0.09}$ & $2.97^{+0.36}_{-0.37}$ & 83/100 \\
2019--2022 & West & $1.33^{+0.36}_{-0.34}$ & $6.64^{+0.03}_{-0.02}$ & $1.74^{+0.10}_{-0.09}$ & $2.41^{+0.39}_{-0.39}$ & 68/84 \\
2019--2022 & North & $0.79^{+0.29}_{-0.27}$ & $6.69^{+0.06}_{-0.03}$ & $1.18^{+0.08}_{-0.07}$ & $3.68^{+0.49}_{-0.49}$ & 40/65 \\
2019--2022 & South & $0.84^{+0.32}_{-0.30}$ & $6.61^{+0.04}_{-0.04}$ & $1.63^{+0.09}_{-0.09}$ & $1.86^{+0.40}_{-0.40}$ & 81/77 \\
\midrule
2023--2025 & All & $5.59^{+0.82}_{-0.81}$ & $6.65^{+0.01}_{-0.02}$ & $6.68^{+0.21}_{-0.21}$ & $2.56^{+0.23}_{-0.23}$ & 133/151 \\
2023--2025 & East & $1.76^{+0.46}_{-0.43}$ & $6.66^{+0.02}_{-0.03}$ & $1.89^{+0.12}_{-0.11}$ & $2.40^{+0.43}_{-0.43}$ & 66/73 \\
2023--2025 & West & $1.45^{+0.45}_{-0.42}$ & $6.65^{+0.03}_{-0.04}$ & $1.96^{+0.12}_{-0.11}$ & $2.43^{+0.43}_{-0.43}$ & 60/73 \\
2023--2025 & North & $1.01^{+0.37}_{-0.34}$ & $6.71^{+0.04}_{-0.03}$ & $1.30^{+0.10}_{-0.09}$ & $2.23^{+0.54}_{-0.53}$ & 39/54 \\
2023--2025 & South & $1.40^{+0.40}_{-0.37}$ & $6.63^{+0.02}_{-0.03}$ & $1.53^{+0.11}_{-0.10}$ & $3.18^{+0.51}_{-0.51}$ & 54/62 \\
\bottomrule
\end{tabular}
\begin{tablenotes}
\item \textit{Notes.}
\item $^\mathrm{a}$ Flux evaluated in the 5.0--8.0 keV band.
\item $^\mathrm{b}$ Uncertainty could not be constrained.
\end{tablenotes}
\end{table*}

\begin{figure*}[ht!]
 \includegraphics[width=1.0\linewidth]{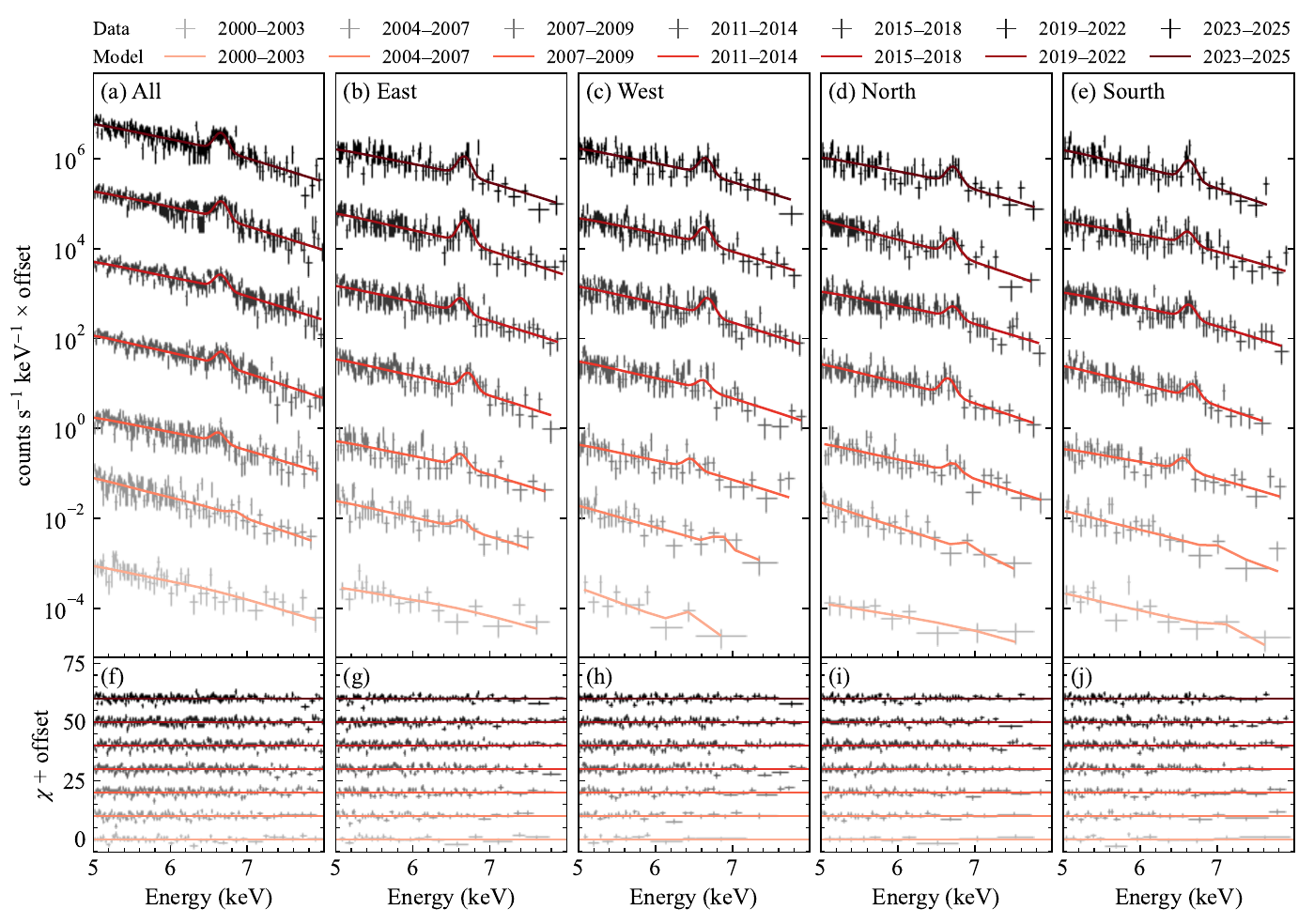}
\caption{Spatially resolved X-ray spectra in the 5.0--8.0 keV band.
Panels from left to right correspond to All, East, West, North, and South.
The top panels (a--e) show the spectra and best-fit models obtained with \texttt{tbabs*(powerlaw+gauss)}, 
while the bottom panels (f--j) show the corresponding residuals.
For clarity, spectra from different epochs are vertically offset in the top panels, 
and the residuals are offset by constant values in the bottom panels. The spectra are ordered by age from bottom (2000--2003) to top (2023--2025). The spectra shown in the top panels are not background subtracted.}
 \label{sn1987a_powerlaw_gauss_data_model}
\end{figure*}

\begin{figure}[ht!]
 \includegraphics[width=1\linewidth]{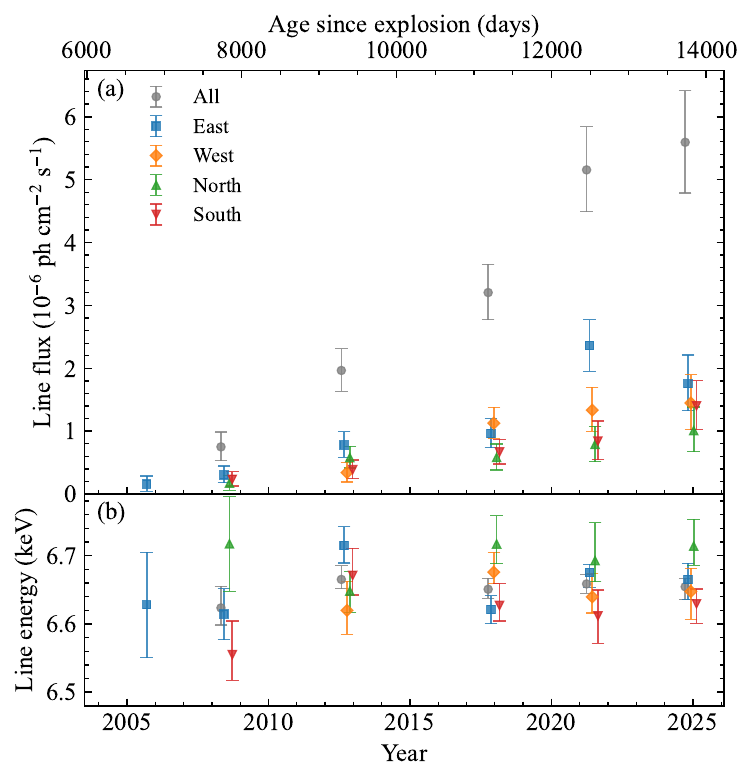}
\caption{Fit results in the 5.0--8.0~keV band obtained with the \texttt{tbabs*(powerlaw + gaussian)} model.
Fits for which the error estimation did not converge are excluded.
(a) Line flux.
(b) Line centroid energy.
The reference ages correspond to the exposure-time-weighted mean ages
listed in Table~\ref{tab:merge_obs_info}; the All region points are
shown at these ages, while those for individual regions are slightly
shifted along the time axis (by offsets of 0.1--0.4~yr) for clarity.}
 \label{sn1987a_gauss_param}
\end{figure}

\begin{figure*}[ht!]
 \includegraphics[width=1.0\linewidth]{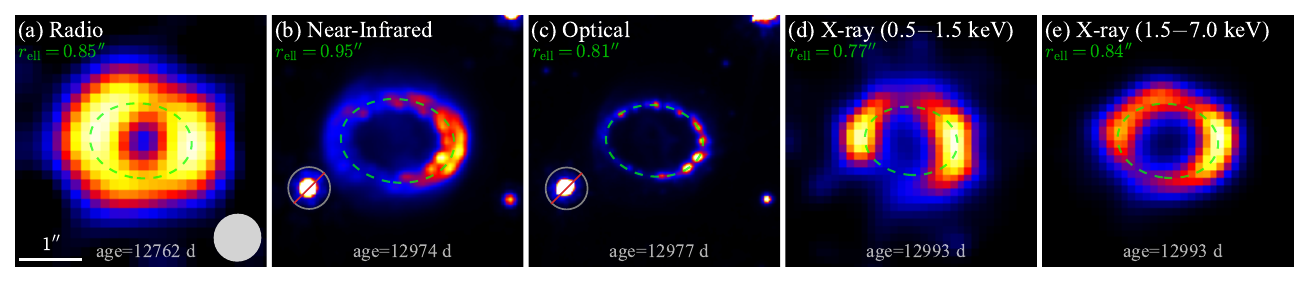}
\caption{Multiwavelength images of SN 1987A obtained in 2022.
(a) Radio (ATCA/9 GHz),
(b) near-infrared (JWST/F444W),
(c) optical (HST/F625W),
(d) X-ray (0.5--1.5 keV), and
(e) X-ray (1.5--7.0 keV).
The X-ray panels are based on a single Chandra observation (ObsID 25514) and processed with 10 iterations of RL deconvolution.
The dashed ellipse indicates the best-fit torus model, and the quoted radius $r_{\mathrm{ell}}$ corresponds to its elliptical radius.
All panels share a common spatial scale.}
 \label{sn1987a_multiwave}
\end{figure*}

\subsection{Spectral analysis}

To investigate the spatial distribution of this Fe~K emission,
we performed spectral analysis for the full region and four sector regions.
Spectral fitting was carried out using an absorbed power-law plus Gaussian model
(\texttt{tbabs*(powerlaw+gauss)}).
Due to limited photon statistics, the 1999--2002 data were excluded from this analysis.
The hydrogen column density was fixed at $N_{\rm H} = 2.17 \times 10^{21}~\mathrm{cm^{-2}}$  \citep{Ravi_2021}. 
The Gaussian width was fixed at $\sigma = 0.04~\mathrm{keV}$,
following the Fe~K line width reported by XRISM \citep{XrismCollaboration_2025}.
The centroid energy of the Gaussian component was allowed to vary within
6.4--7.0 keV.
Fits yielding centroid energies outside the range 6.5--6.8 keV were discarded as they are unlikely to correspond to the Fe~K emission \citep[e.g.,][]{Sun_2021}.
The best-fit models are shown in Figure~\ref{sn1987a_powerlaw_gauss_data_model}
and summarized in Table~\ref{tab:powerlaw_gauss}.

The temporal evolution of the Gaussian parameters is shown in Figure~\ref{sn1987a_gauss_param}.
For the All region, the Fe~K line emission becomes visible in the
2007--2009 epoch and shows a general increasing trend toward later epochs.
The line flux and centroid energy are broadly consistent with those reported
in XMM-Newton studies \citep{Sun_2021, Sun_2025}.

For the spatially resolved results, previous Chandra studies have suggested
that the eastern region becomes systematically brighter in Fe~K emission
after $\sim$2018 \citep{Ravi_2024}.
Our results are consistent with this trend and further suggest that this asymmetry
may already be present in earlier epochs.
In particular, the eastern region shows the earliest emergence of the Fe~K line
in the 2004--2007 epoch and exhibits a systematically higher line intensity
throughout the subsequent epochs.
A similar trend is seen in the continuum component, where the eastern region is brighter at earlier epochs (Table~\ref{tab:powerlaw_gauss}).

\subsection{Multiwavelength Morphology in 2022}

Multiwavelength imaging of SN~1987A has been presented at
various epochs (e.g., \citealt{Burrows_2000};
\citealt{Frank_2016}). Here, we show the 2022 observations
as an updated snapshot of the remnant.

Figure~\ref{sn1987a_multiwave} displays radio (ATCA/9~GHz),
near-infrared (JWST/F444W), optical (HST/F625W), and X-ray
(0.5--1.5~keV and 1.5--7.0~keV) images on a common spatial scale.
For reference, we applied the same torus model described in Appendix~\ref{app:torus_model} to each image in order to characterize the ring structure.
The derived radii are
$r_{\mathrm{ell}} = 0\farcs850\pm0\farcs006$ (radio),
$0\farcs949\pm0\farcs003$ (near-infrared),
$0\farcs811\pm0\farcs007$ (optical),
$0\farcs774\pm0\farcs046$ (0.5--1.5~keV), and
$0\farcs841\pm0\farcs011$ (1.5--7.0~keV).
These values provide a useful reference for comparing the radial structure across wavelengths, 
although care is required in their interpretation, as different modeling approaches are commonly adopted across wavelengths \citep[e.g.,][]{Ng_2009, Ng_2013}, 
and the derived radii can also be influenced by instrumental effects and observational configurations \citep[e.g.,][]{Ravi_2024}.

As a reference for the multiwavelength comparison, we briefly describe the radio morphology in 2022. 
Compared to earlier radio studies, in which the eastern side was consistently brighter than the western side in earlier epochs \citep[e.g.,][]{Ng_2013,Cendes_2018},
the radio image in 2022 appears more azimuthally symmetric, with the eastern and western intensities becoming comparable. 
In contrast, the images at other wavelengths in 2022 show brighter emission on the western side. 

\section{Discussion and Summary}

In this study, we investigated the temporal evolution of the X-ray
morphology and Fe~K emission of SN~1987A using a uniformly analyzed
long-term Chandra dataset. By combining long-term imaging and spectral
analysis, we examined the energy-dependent evolution of the remnant
structure and its relation to the changing physical conditions within
SN~1987A. The main findings and their possible interpretations are
summarized below.

First, the ring width shows an energy dependence after the early 2010s,
with the soft X-ray band (0.5--1.5~keV) becoming systematically broader
than the hard band (1.5--7.0~keV). Importantly, this difference is not
simply a radial offset between the two bands. By considering the radius
together with the width, we find that the soft component extends inward
at later epochs. In particular, the inner boundary of the soft-band
distribution reaches $r_{\mathrm{ell}} \sim 0\farcs5$ in the 2020s,
comparable to the radius observed around the year 2000. Previous studies
showed a rapid increase in the soft X-ray flux around $\sim$6000--7000
days, likely due to the interaction of the blast wave with dense clumps
in the circumstellar medium \citep{Frank_2016}. This interpretation is
further supported by recent JWST observations, which show that the
reverse shock extends from just inside the ER to higher latitudes and
forms a spatially extended, bubble-like structure \citep{Larsson_2023}.
Within this picture, the inward extension of the soft X-ray emission may
reflect the increasing contribution of emission from
reverse-shock-heated material and/or higher-latitude structures within
the remnant. This interpretation is consistent with recent HD/MHD
simulations, which predict an increasing contribution from
reverse-shock-heated ejecta, particularly in the soft X-ray band
\citep{Orlando_2015, Orlando_2019, Orlando_2020, Orlando_2025}.

Second, the Fe~K emission detected in our analysis is broadly consistent
with previous XMM-Newton and XRISM studies
\citep{Sun_2021, Sun_2025, XrismCollaboration_2025}. While this does not
represent a new discovery of Fe~K emission from SN~1987A, the present
analysis provides an independent Chandra-based confirmation using a
uniformly reduced long-term dataset. The spatially resolved analysis
further suggests that the Fe~K emission tends to be stronger in the
eastern region, consistent with the trend reported by \citet{Ravi_2024}.
Although this trend was reported most clearly after $\sim$2018 by
\citet{Ravi_2024}, our results suggest that a similar tendency may
already be present at earlier epochs, although the statistical
significance in the earlier data remains limited.

Finally, the radial ordering of soft X-ray, optical, and hard X-ray
emission in 2022 suggests that the soft component arises from regions
interior to the optical ring, likely reflecting the interaction of the
shock with the ER. Previous studies have also suggested that the
east--west asymmetry in the hard X-ray band begins to reverse around
9500 days \citep{Frank_2016}, while the radio emission evolves in a
similar manner but with a delay of $\sim$2000 days \citep{Cendes_2018}.
The present radio observation appears broadly consistent with this delayed evolution,
which is likely related to magnetic field amplification following the
passage of the shock \citep[e.g.,][]{Cendes_2018}. Overall, our results
indicate that SN~1987A is transitioning from an ER-dominated phase to
one in which emission from interior components becomes increasingly
important.

\begin{acknowledgments}
We thank the anonymous referee for the constructive and thoughtful comments that helped improve this work.
X-ray image analysis was performed using CIAO, and subpixel PSF simulations were carried out with MARX.
Spectral fitting was performed using XSPEC, and radio data reduction was conducted with CASA.
This paper includes data obtained with the Australia Telescope Compact Array (ATCA), which is part of the Australia Telescope National Facility funded by the Australian Government for operation as a National Facility managed by CSIRO. This research has made use of the Mikulski Archive for Space Telescopes (MAST). We acknowledge the developers and calibration teams of these tools for their indispensable contributions.
We thank Yoshiaki Kanemaru, Takaaki Tanaka, and Hideki Uchiyama for the helpful discussions and comments.
This work was supported by JSPS KAKENHI Grant Nos. JP24KJ2067(Y.S.), 22H01272(S.Y.), 23K22543(S.Y.), 24K00672(S.Y.), 23K20850(K.M.), 22K20386(H.S.), 23K13148(H.S.), 26K17195(H.S.), and 24KJ1485(K.M.).
\end{acknowledgments}

\section*{Data Availability}

This paper employs a list of Chandra datasets, obtained by the Chandra X-ray Observatory, contained in the Chandra Data Collection~\dataset[doi: 10.25574/cdc.594]{https://doi.org/10.25574/cdc.594}.
The HST and JWST data presented in this article were obtained from the Mikulski Archive for Space Telescopes (MAST) at the Space Telescope Science Institute. The specific observations analyzed can be accessed via \dataset[doi: 10.17909/6ejt-dd11]{https://doi.org/10.17909/6ejt-dd11}.




\appendix

\section{Observation Log and Data Summary}
\label{app:obslog}

\begin{figure*}[ht!]
 \includegraphics[width=1.0\linewidth]{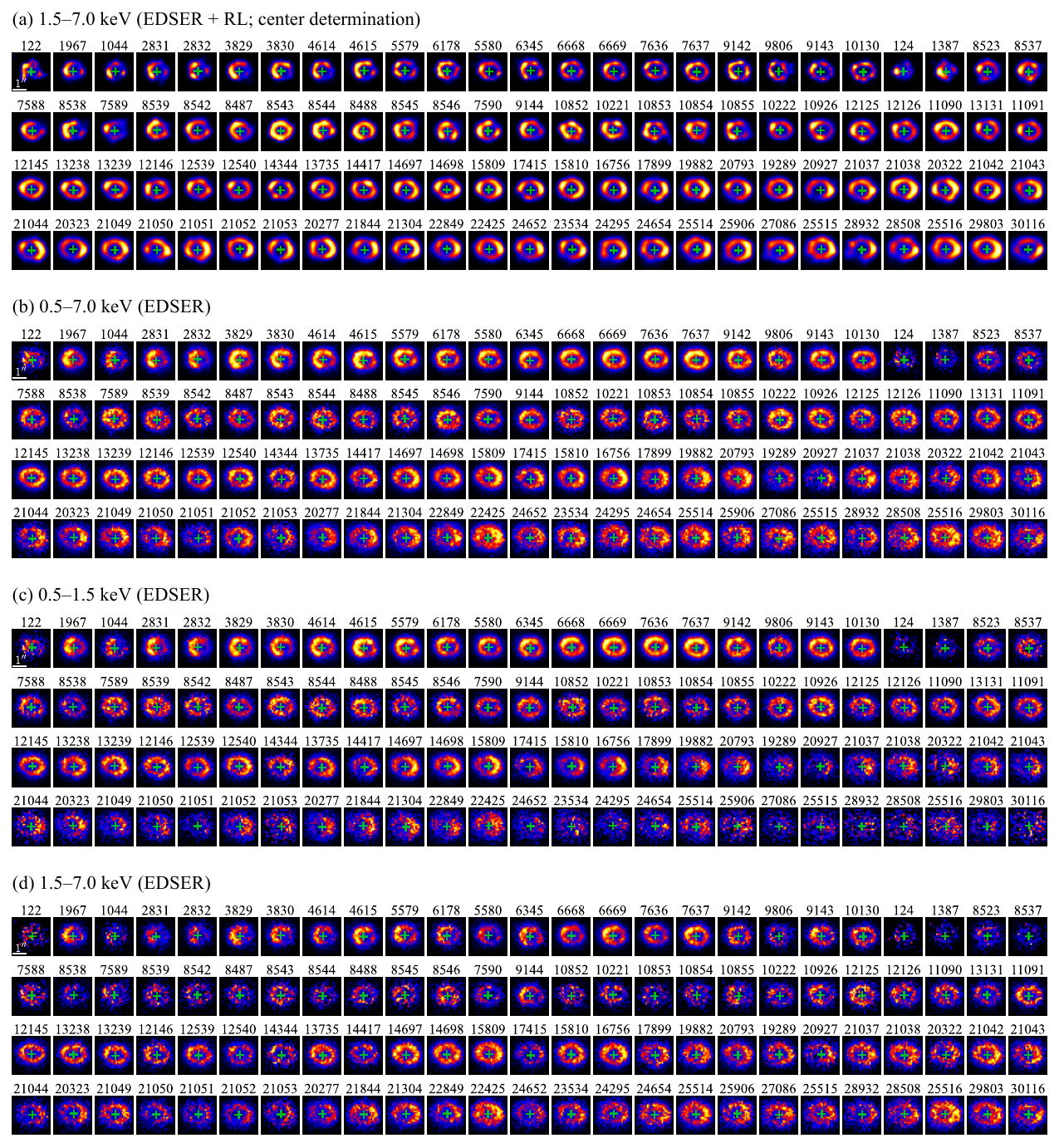}
\caption{Center determination based on Chandra/ACIS-S observations (NONE and HETG).
(a) RL-deconvolved (10 iterations) 1.5--7.0~keV images reconstructed from EDSER-processed data (1/4-pixel resolution), used for torus-model fitting to determine the center position.
(b) Observed 0.5--7.0~keV images.
(c) Observed 0.5--1.5~keV images.
(d) Observed 1.5--7.0~keV images.
The ``+'' symbols indicate the derived center positions, which are adopted as the reference for image alignment and merging.
The corresponding ObsID is shown in each panel.
The color scale in each panel is independently normalized.}
\label{sn1987a_center_image_evaluation}
\end{figure*}

This appendix summarizes the Chandra/ACIS-S observations of SN~1987A used in this work.
Table~\ref{tab:obslog} lists the observational parameters and the derived center coordinates.
The center positions were obtained via torus model fitting to the 1.5--7.0~keV images and were used as the reference for image alignment and merging.
Representative images with the derived center positions overlaid are shown in Figure~\ref{sn1987a_center_image_evaluation}, demonstrating the consistency and robustness of the center determination.

\begin{longtable}{r l r r r c c l l}
\caption{Chandra observations of SN~1987A used in this paper.} \label{tab:obslog} \\
\toprule
\toprule
ObsID &
\multicolumn{1}{c}{Start Date} &
\multicolumn{1}{c}{Age} &
\multicolumn{1}{c}{Exposure} &
\multicolumn{1}{c}{Counts$^*$} &
\multicolumn{1}{c}{Frame Time} &
Grating &
\multicolumn{1}{c}{Center RA$^\dagger$} &
\multicolumn{1}{c}{Center Dec$^\dagger$} \\
 &
\multicolumn{1}{c}{(yyyy mmm dd)} &
\multicolumn{1}{c}{(days)} &
\multicolumn{1}{c}{(ks)} &
 &
\multicolumn{1}{c}{(s)} &
 &
\multicolumn{1}{c}{(deg)} &
\multicolumn{1}{c}{(deg)} \\
\midrule
\midrule
\endfirsthead
\caption[]{Chandra observations of SN~1987A used in this paper.} \\
\toprule
\toprule
ObsID &
\multicolumn{1}{c}{Start Date} &
\multicolumn{1}{c}{Age} &
\multicolumn{1}{c}{Exposure} &
\multicolumn{1}{c}{Counts} &
\multicolumn{1}{c}{Frame Time} &
Grating &
\multicolumn{1}{c}{Center RA} &
\multicolumn{1}{c}{Center Dec} \\
 &
\multicolumn{1}{c}{(yyyy mmm dd)} &
\multicolumn{1}{c}{(days)} &
\multicolumn{1}{c}{(ks)} &
 &
\multicolumn{1}{c}{(s)} &
 &
\multicolumn{1}{c}{(deg)} &
\multicolumn{1}{c}{(deg)} \\
\midrule
\midrule
\endhead
\midrule
\multicolumn{9}{r}{Continued on next page} \\
\midrule
\endfoot
\bottomrule
\multicolumn{9}{l}{
\parbox{0.95\textwidth}{
\textit{Notes.} \\
$^*$ Counts are calculated in the 0.5--7.0 keV band within a circular aperture of radius $3''$ centered on the torus-fitted source position. \\
$^\dagger$ The center coordinates are derived independently for each observation from torus fitting using the native WCS of the corresponding dataset. Values in parentheses indicate the uncertainty in the last quoted digit.
}
}
\endlastfoot
122 & 2000 Jan 17 & 4711 & 8.6 & 592 & 3.2 & NONE & 83.866601(77) & -69.269738(28) \\
1967 & 2000 Dec 7 & 5036 & 98.8 & 8710 & 3.2 & NONE & 83.866623(11) & -69.269780(4) \\
1044 & 2001 Apr 25 & 5175 & 17.8 & 1736 & 3.2 & NONE & 83.866605(22) & -69.269736(14) \\
2831 & 2001 Dec 12 & 5406 & 49.4 & 5987 & 3.1 & NONE & 83.866789(14) & -69.269797(6) \\
2832 & 2002 May 15 & 5560 & 44.3 & 6200 & 3.1 & NONE & 83.866547(15) & -69.269738(6) \\
3829 & 2002 Dec 31 & 5790 & 49.0 & 8909 & 3.1 & NONE & 83.866661(13) & -69.269739(5) \\
3830 & 2003 Jul 8 & 5979 & 45.3 & 9333 & 3.1 & NONE & 83.866763(13) & -69.269748(5) \\
4614 & 2004 Jan 2 & 6157 & 46.5 & 11475 & 3.1 & NONE & 83.866670(11) & -69.269767(3) \\
4615 & 2004 Jul 22 & 6359 & 48.8 & 17427 & 1.5 & NONE & 83.866545(8) & -69.269714(3) \\
5579 & 2005 Jan 9 & 6530 & 31.9 & 15813 & 0.4 & NONE & 83.866752(9) & -69.269792(3) \\
6178 & 2005 Jan 13 & 6534 & 16.5 & 8300 & 0.4 & NONE & 83.866731(13) & -69.269793(4) \\
5580 & 2005 Jul 11 & 6714 & 23.8 & 14121 & 0.4 & NONE & 83.866580(9) & -69.269711(3) \\
6345 & 2005 Jul 16 & 6718 & 21.0 & 12436 & 0.4 & NONE & 83.866542(11) & -69.269724(4) \\
6668 & 2006 Jan 28 & 6914 & 42.3 & 30014 & 0.4 & NONE & 83.866664(7) & -69.269785(2) \\
6669 & 2006 Jul 27 & 7095 & 36.5 & 30132 & 0.4 & NONE & 83.866634(7) & -69.269711(2) \\
7636 & 2007 Jan 19 & 7271 & 33.5 & 31767 & 0.4 & NONE & 83.866715(6) & -69.269753(2) \\
7637 & 2007 Jul 13 & 7445 & 25.7 & 27155 & 0.4 & NONE & 83.866545(7) & -69.269727(2) \\
9142 & 2008 Jan 9 & 7625 & 6.6 & 8209 & 0.2 & NONE & 83.866826(13) & -69.269747(5) \\
9806 & 2008 Jan 11 & 7627 & 2.6 & 3458 & 0.2 & NONE & 83.866729(30) & -69.269761(8) \\
9143 & 2008 Jul 4 & 7802 & 8.6 & 11768 & 0.2 & NONE & 83.866571(11) & -69.269718(3) \\
10130 & 2009 Jan 5 & 7987 & 6.0 & 8945 & 0.2 & NONE & 83.866902(10) & -69.269789(3) \\
124 & 1999 Oct 6 & 4608 & 47.3 & 282 & 3.2 & HETG & 83.866901(61) & -69.269837(18) \\
1387 & 1999 Oct 6 & 4608 & 68.9 & 403 & 3.2 & HETG & 83.866978(60) & -69.269823(17) \\
8523 & 2007 Mar 11 & 7321 & 29.6 & 2279 & 2.5 & HETG & 83.866666(18) & -69.269787(7) \\
8537 & 2007 Mar 12 & 7322 & 12.7 & 1062 & 2.5 & HETG & 83.866712(31) & -69.269733(9) \\
7588 & 2007 Mar 13 & 7324 & 27.2 & 2079 & 2.5 & HETG & 83.866702(18) & -69.269795(6) \\
8538 & 2007 Mar 18 & 7328 & 20.7 & 1786 & 2.5 & HETG & 83.866640(27) & -69.269805(7) \\
7589 & 2007 Mar 19 & 7329 & 25.3 & 2064 & 2.5 & HETG & 83.866595(24) & -69.269757(7) \\
8539 & 2007 Mar 20 & 7330 & 24.8 & 2014 & 2.5 & HETG & 83.866621(20) & -69.269793(6) \\
8542 & 2007 Mar 21 & 7331 & 17.9 & 1401 & 2.5 & HETG & 83.866694(22) & -69.269798(8) \\
8487 & 2007 Mar 24 & 7334 & 28.7 & 2222 & 2.5 & HETG & 83.866662(21) & -69.269807(7) \\
8543 & 2007 Mar 27 & 7337 & 30.7 & 2527 & 2.5 & HETG & 83.866608(16) & -69.269812(5) \\
8544 & 2007 Mar 28 & 7338 & 19.1 & 1610 & 2.5 & HETG & 83.866647(22) & -69.269782(7) \\
8488 & 2007 Mar 29 & 7339 & 31.7 & 2649 & 2.5 & HETG & 83.866643(15) & -69.269791(5) \\
8545 & 2007 Mar 31 & 7341 & 20.5 & 1717 & 2.5 & HETG & 83.866594(21) & -69.269793(7) \\
8546 & 2007 Apr 1 & 7342 & 30.6 & 2674 & 2.5 & HETG & 83.866626(19) & -69.269788(6) \\
7590 & 2007 Apr 17 & 7358 & 35.5 & 2953 & 2.5 & HETG & 83.866621(13) & -69.269790(5) \\
9144 & 2008 Jul 1 & 7799 & 42.0 & 5073 & 1.1 & HETG & 83.866536(13) & -69.269726(4) \\
10852 & 2009 Jan 12 & 7994 & 10.8 & 1427 & 1.1 & HETG & 83.866519(22) & -69.269745(7) \\
10221 & 2009 Jan 13 & 7995 & 18.7 & 2393 & 1.1 & HETG & 83.866665(17) & -69.269771(5) \\
10853 & 2009 Jan 15 & 7997 & 11.2 & 1438 & 1.1 & HETG & 83.866581(24) & -69.269730(7) \\
10854 & 2009 Jan 17 & 7999 & 12.0 & 1590 & 1.1 & HETG & 83.866596(23) & -69.269755(8) \\
10855 & 2009 Jan 18 & 8000 & 18.8 & 2462 & 1.1 & HETG & 83.866627(23) & -69.269763(6) \\
10222 & 2009 Jul 6 & 8170 & 24.4 & 3487 & 1.1 & HETG & 83.866406(12) & -69.269707(4) \\
10926 & 2009 Sep 8 & 8233 & 33.8 & 4915 & 1.1 & HETG & 83.866729(14) & -69.269717(4) \\
12125 & 2010 Mar 17 & 8423 & 18.1 & 2635 & 1.0 & HETG & 83.866537(19) & -69.269780(5) \\
12126 & 2010 Mar 17 & 8423 & 21.2 & 3069 & 1.0 & HETG & 83.866481(15) & -69.269750(5) \\
11090 & 2010 Mar 28 & 8434 & 24.6 & 3529 & 1.0 & HETG & 83.866541(14) & -69.269768(4) \\
13131 & 2010 Sep 28 & 8618 & 26.5 & 3981 & 1.0 & HETG & 83.866678(13) & -69.269700(4) \\
11091 & 2010 Sep 29 & 8619 & 27.9 & 4267 & 1.0 & HETG & 83.866675(12) & -69.269777(3) \\
12145 & 2011 Mar 1 & 8772 & 51.2 & 7540 & 2.5 & HETG & 83.866611(9) & -69.269804(3) \\
13238 & 2011 Mar 4 & 8775 & 54.2 & 7972 & 2.5 & HETG & 83.866582(9) & -69.269802(2) \\
13239 & 2011 Mar 6 & 8777 & 46.7 & 7124 & 2.5 & HETG & 83.866544(10) & -69.269764(3) \\
12146 & 2011 Mar 13 & 8784 & 25.6 & 3997 & 2.5 & HETG & 83.866580(14) & -69.269808(4) \\
12539 & 2011 Mar 25 & 8796 & 52.1 & 8196 & 1.0 & HETG & 83.866476(8) & -69.269758(3) \\
12540 & 2011 Sep 21 & 8976 & 37.5 & 6132 & 1.0 & HETG & 83.866545(10) & -69.269664(3) \\
14344 & 2011 Sep 22 & 8978 & 11.6 & 1838 & 1.0 & HETG & 83.866501(15) & -69.269682(5) \\
13735 & 2012 Mar 28 & 9166 & 42.9 & 7271 & 1.0 & HETG & 83.866639(10) & -69.269814(3) \\
14417 & 2012 Apr 1 & 9169 & 26.9 & 4421 & 1.0 & HETG & 83.866643(11) & -69.269761(3) \\
14697 & 2013 Mar 21 & 9524 & 67.6 & 11689 & 1.1 & HETG & 83.866702(6) & -69.269871(2) \\
14698 & 2013 Sep 28 & 9714 & 68.5 & 11997 & 1.1 & HETG & 83.866728(7) & -69.269651(2) \\
15809 & 2014 Mar 19 & 9886 & 70.5 & 11984 & 1.1 & HETG & 83.866413(7) & -69.269836(2) \\
17415 & 2014 Sep 17 & 10069 & 19.4 & 3175 & 1.1 & HETG & 83.866695(13) & -69.269758(4) \\
15810 & 2014 Sep 20 & 10071 & 48.3 & 8069 & 1.1 & HETG & 83.866755(7) & -69.269748(2) \\
16756 & 2015 Sep 17 & 10433 & 66.6 & 10484 & 1.1 & HETG & 83.866472(7) & -69.269677(2) \\
17899 & 2016 Sep 19 & 10801 & 26.1 & 3893 & 1.1 & HETG & 83.866498(13) & -69.269765(4) \\
19882 & 2016 Sep 23 & 10805 & 41.1 & 6202 & 1.1 & HETG & 83.866651(10) & -69.269776(3) \\
20793 & 2017 Sep 21 & 11169 & 48.3 & 6601 & 1.1 & HETG & 83.866626(9) & -69.269726(3) \\
19289 & 2017 Sep 23 & 11170 & 18.9 & 2556 & 1.1 & HETG & 83.866747(14) & -69.269782(4) \\
20927 & 2018 Mar 14 & 11342 & 16.6 & 2038 & 1.7 & HETG & 83.866369(17) & -69.269798(5) \\
21037 & 2018 Mar 15 & 11343 & 29.4 & 3809 & 1.7 & HETG & 83.866313(10) & -69.269893(3) \\
21038 & 2018 Mar 18 & 11347 & 34.2 & 4240 & 1.7 & HETG & 83.866399(11) & -69.269886(3) \\
20322 & 2018 Mar 19 & 11348 & 15.2 & 1840 & 1.7 & HETG & 83.866307(17) & -69.269765(5) \\
21042 & 2018 Mar 23 & 11351 & 41.0 & 5250 & 1.7 & HETG & 83.866350(9) & -69.269883(3) \\
21043 & 2018 Mar 25 & 11353 & 28.8 & 3603 & 1.7 & HETG & 83.866356(12) & -69.269825(4) \\
21044 & 2018 Mar 26 & 11354 & 15.2 & 1910 & 1.7 & HETG & 83.866333(17) & -69.269836(5) \\
20323 & 2018 Mar 27 & 11355 & 27.1 & 3333 & 1.7 & HETG & 83.866398(12) & -69.269781(4) \\
21049 & 2018 Mar 28 & 11356 & 30.9 & 3866 & 1.7 & HETG & 83.866332(12) & -69.269821(4) \\
21050 & 2018 Mar 29 & 11357 & 17.8 & 2187 & 1.7 & HETG & 83.866334(15) & -69.269767(5) \\
21051 & 2018 Mar 30 & 11358 & 15.0 & 1935 & 1.7 & HETG & 83.866344(19) & -69.269762(6) \\
21052 & 2018 Mar 31 & 11359 & 30.1 & 3677 & 1.7 & HETG & 83.866422(9) & -69.269876(4) \\
21053 & 2018 Apr 2 & 11361 & 12.2 & 1572 & 1.7 & HETG & 83.866348(17) & -69.269753(6) \\
20277 & 2018 Sep 15 & 11527 & 33.8 & 4426 & 1.1 & HETG & 83.866755(12) & -69.269766(3) \\
21844 & 2018 Sep 16 & 11528 & 33.8 & 4397 & 1.1 & HETG & 83.866826(9) & -69.269758(4) \\
21304 & 2019 Sep 17 & 11894 & 41.1 & 4818 & 1.1 & HETG & 83.866585(11) & -69.269664(3) \\
22849 & 2019 Sep 18 & 11895 & 41.1 & 4813 & 1.1 & HETG & 83.866635(11) & -69.269680(3) \\
22425 & 2020 Sep 12 & 12255 & 60.8 & 6835 & 1.1 & HETG & 83.866318(9) & -69.269716(3) \\
24652 & 2020 Sep 17 & 12260 & 29.3 & 3310 & 1.1 & HETG & 83.866683(11) & -69.269875(4) \\
23534 & 2021 Oct 25 & 12664 & 27.1 & 2890 & 1.1 & HETG & 83.866648(10) & -69.269988(4) \\
24295 & 2021 Oct 27 & 12665 & 29.0 & 3018 & 1.1 & HETG & 83.866698(15) & -69.269930(4) \\
24654 & 2021 Oct 28 & 12666 & 31.9 & 3289 & 1.1 & HETG & 83.866748(12) & -69.270020(4) \\
25514 & 2022 Sep 21 & 12994 & 57.0 & 5809 & 1.1 & HETG & 83.866928(9) & -69.269801(3) \\
25906 & 2022 Sep 27 & 13000 & 26.4 & 2685 & 1.1 & HETG & 83.867138(16) & -69.269929(4) \\
27086 & 2023 Sep 12 & 13350 & 28.1 & 2685 & 1.1 & HETG & 83.866666(18) & -69.269686(4) \\
25515 & 2023 Sep 26 & 13364 & 35.3 & 3262 & 1.1 & HETG & 83.866849(15) & -69.269857(4) \\
28932 & 2023 Sep 27 & 13365 & 21.3 & 1974 & 1.1 & HETG & 83.866755(19) & -69.269859(6) \\
28508 & 2024 Sep 27 & 13731 & 28.8 & 2550 & 1.1 & HETG & 83.866904(17) & -69.269719(5) \\
25516 & 2024 Oct 1 & 13735 & 55.9 & 5083 & 1.1 & HETG & 83.866729(12) & -69.269905(3) \\
29803 & 2025 Sep 17 & 14086 & 54.1 & 4818 & 1.1 & HETG & 83.866542(11) & -69.269830(3) \\
30116 & 2025 Oct 2 & 14101 & 29.9 & 2680 & 1.0 & HETG & 83.866705(18) & -69.269776(4) \\
\end{longtable}

\section{Elliptical Gaussian Torus Model}
\label{app:torus_model}

The ER of SN~1987A is intrinsically close to circular but appears elliptical in projection due to its inclination with respect to the line of sight. We therefore model the observed emission using an elliptical Gaussian torus profile defined in the projected image plane.

The image coordinates $(x, y)$ are shifted to a coordinate system centered at $(x_0, y_0)$ 
and rotated by the position angle $\mathrm{PA}$:
\begin{equation}
\begin{aligned}
x' &= (x-x_0)\cos(\mathrm{PA}) + (y-y_0)\sin(\mathrm{PA}), \\
y' &= -(x-x_0)\sin(\mathrm{PA}) + (y-y_0)\cos(\mathrm{PA}),
\end{aligned}
\end{equation}
where $(x', y')$ denote the rotated coordinates.

The elliptical radial coordinate accounting for the inclination is defined as
\begin{equation}
r_{\mathrm{ell}} =
\sqrt{
x'^2 + \left(\frac{y'}{\cos i}\right)^2
},
\end{equation}
which corresponds to the projected radius of a circular ring viewed at an inclination angle $i$.

The surface brightness distribution is modeled as a Gaussian ring
peaking at $r_{\mathrm{ell}} = r$:
\begin{equation}
I(x,y) = N_r \exp\left(
-\frac{(r_{\mathrm{ell}} - r)^2}{2\sigma_r^2}
\right) + N_0,
\end{equation}
where $r$ is the ring radius, $\sigma_r$ is the radial width,
$N_r$ is the ring normalization, and $N_0$ represents a uniform background.
The inclination angle $i$ and position angle $\mathrm{PA}$ are fixed to 
$i = 42\fdg85$ and $\mathrm{PA} = -6\fdg24$, based on optical measurements 
of the ER geometry with HST \citep{Tegkelidis_2024}, while the remaining 
parameters are fitted.

The free parameters of the model are $(x_0, y_0, r, \sigma_r, N_r, N_0)$.
For the 4 yr merged datasets, $(x_0, y_0)$ was fixed, as it had
already been determined independently. For the X-ray images and
multiwavelength datasets, a standard $\chi^2$ statistic was adopted for
the torus-model fitting. The statistical uncertainties of the fitted
parameters for the X-ray images were estimated by Monte Carlo
simulations. For each fitted X-ray image, including both merged
datasets and individual observations, we generated 300 Poisson
realizations of the observed counts image, applied RL deconvolution and
torus-model fitting procedures, and derived the parameter uncertainties
from the standard deviations of the resulting best-fit parameter
distributions.

\section{Torus Fitting Results for Individual Observations}
\label{appendix:individual_torus}

\begin{figure}[ht!]
 \includegraphics[width=1.0\linewidth]{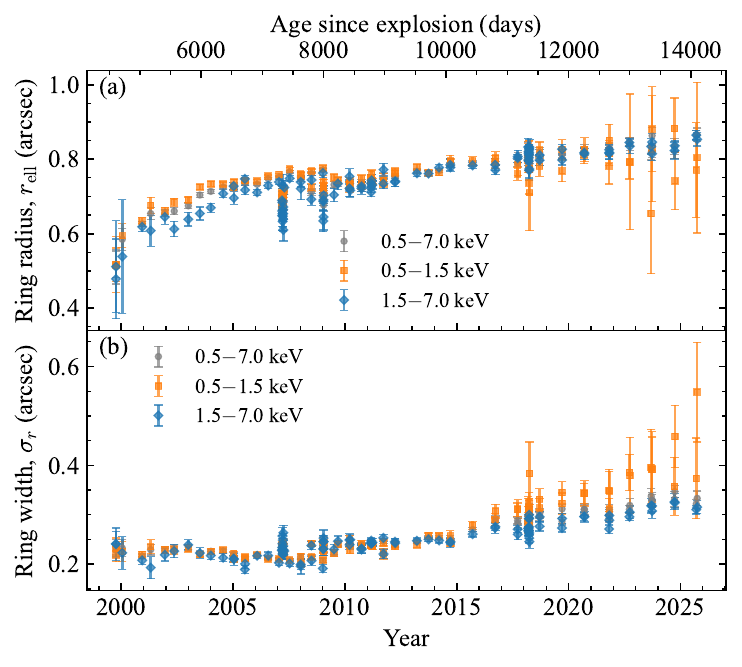}
\caption{Temporal evolution of the torus radius ($r_{\mathrm{ell}}$) and width ($\sigma_r$) derived from RL-deconvolved images for individual observations. Results are shown for the 0.5--7.0 (gray), 0.5--1.5 (orange), and 1.5--7.0~keV (blue) bands.
(a) Torus radius $r_{\mathrm{ell}}$.
(b) Torus width $\sigma_r$.}
 \label{sn1987a_torus_param_bin_0.25_eachobs}
\end{figure}

To evaluate the robustness of the trends derived from the 4 yr
merged images, we additionally performed torus fitting for individual
observations. For this analysis, RL deconvolution was performed with
10 iterations in order to suppress excessive noise amplification in the
lower-count datasets. Figure~\ref{sn1987a_torus_param_bin_0.25_eachobs}
shows the temporal evolution of the torus radius and width measured
from the RL-deconvolved images for each observation.

Although the scatter is naturally larger than in the merged analysis
because of the lower photon statistics of individual observations, the
overall trends remain consistent with the results derived from the
merged datasets. In particular, the soft-band emission tends to show a
larger torus width than the hard band after the early 2010s, consistent
with the main results presented in
Figure~\ref{sn1987a_torus_param_bin_0.25_yearbin_4}. These results
indicate that the observed energy dependence of the torus width is not
driven solely by the merging procedure and is already present in the
individual observations.

\section{Spatially Resolved Radial Profile Analysis}
\label{appendix:radial_profile}

We evaluate the width of the torus structure using a data-driven approach
based on radial profiles, complementing the model fitting results presented
in Section~\ref{sec:imaging}. The radial profiles were constructed in the
sky plane using circular annuli without deprojection or elliptical
coordinate transformation. The center positions of the annuli were taken
from the torus fitting results for the corresponding images. We adopted
circular annuli in order to provide a complementary measure with reduced
model dependence compared with elliptical or deprojected analyses.

\begin{figure*}[ht!]
 \includegraphics[width=1.0\linewidth]{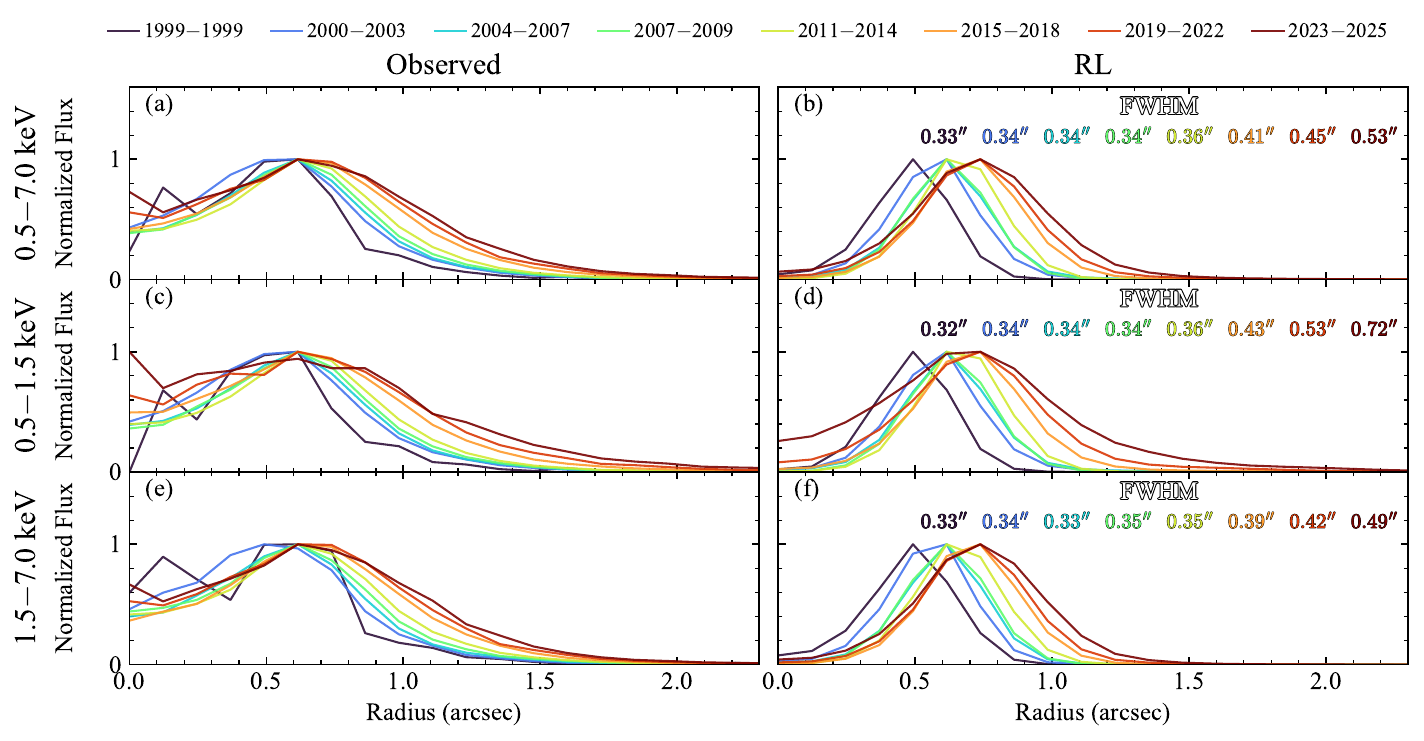}
\caption{Radial profiles of the 4 yr binned X-ray images shown in Figure~\ref{sn1987a_torus_image_obs_rl_iter_30_bin_0.25_yearbin_4}.
The left panels show profiles from the EDSER-reconstructed (1/4 pixel) images, while the right panels show those from RL-deconvolved images.
From top to bottom, the panels correspond to the 0.5--7.0, 0.5--1.5, and 1.5--7.0~keV bands.
All profiles are normalized to their respective peak values and are color-coded by epoch, with the FWHM values derived from the RL profiles indicated in the corresponding colors.}
 \label{sn1987a_radial_profile_yearbin_4}
\end{figure*}

The radial profiles were constructed from the 4 yr binned images
(Figure~\ref{sn1987a_torus_image_obs_rl_iter_30_bin_0.25_yearbin_4})
using the same center positions and are shown in
Figure~\ref{sn1987a_radial_profile_yearbin_4}. Both the observed and
RL-deconvolved profiles show an outward shift with time, indicating overall
expansion, and the soft band exhibits more extended outer wings than the
hard band.

We quantify these trends using the full width at half-maximum (FWHM),
primarily measured from the RL-deconvolved profiles. The results agree
with the width evolution derived from the torus model
(Figure~\ref{sn1987a_torus_param_bin_0.25_yearbin_4}), showing an increasing
trend with time in the 0.5--7.0~keV band, while the FWHM in the hard band
(1.5--7.0 keV) is smaller than that in the soft band (0.5--1.5 keV) after
2015--2018. We also examined the FWHM in individual spatial regions
(East, West, North, and South; $\pm45^\circ$ sectors), which show similar
energy-dependent trends. A detailed analysis of the spatial variation is
beyond the scope of this work. Overall, these results follow the trends
reported in Section~\ref{sec:imaging}.

\bibliography{main}{}
\bibliographystyle{aasjournalv7}



\end{document}